\pdfoutput=1
\documentclass{article}
\usepackage[utf8]{inputenc}

\usepackage{amsmath,amsthm,amssymb}
\usepackage[dvipsnames]{xcolor}
\usepackage{physics}
\usepackage{tensor}
\usepackage{graphicx}
\usepackage{float}
\usepackage{xcolor, colortbl}
\usepackage{hyperref}
\usepackage{authblk}
\usepackage{mathtools}

\definecolor{light-gray}{gray}{0.9} 

\usepackage{setspace}
\newcommand{\sltr}{\mathrm{SL}(2,\mathbb{R})}
\newcommand{\sltra}{\mathfrak{sl}(2,\mathbb{R})}

\newenvironment{eqaed}
    {
    \begin{equation}
        \begin{aligned}
    }
    {
        \end{aligned}
    \end{equation}
    }

\title{On self-similar axion-dilaton configurations}
\author[1]{Riccardo Antonelli}
\author[1]{Ehsan Hatefi}

\affil[1]{Scuola Normale Superiore and I.N.F.N,\protect\\ 
Piazza dei Cavalieri 7, 56126, Pisa, Italy\vspace{1em}}
\affil[ ]{\small\textit{riccardo.antonelli@sns.it, ehsan.hatefi@sns.it}}

\begin{document}

\maketitle

\begin{abstract}
    
    
    \noindent We explore self-similar collapse solutions to the Einstein-axion-dilaton system in arbitrary spacetime dimensions, which are invariant under spacetime dilation combined with internal $\sltr$ transformations. We set up a numerical formalism, and test it in four and five dimensions, discovering several new families of solutions in the three conjugacy classes of $\sltra$.
\end{abstract}

\section{Introduction}

A curious gedanken experiment in gravitation, performed in~\cite{Chop}, explored the interface between the initial conditions that lead to collapse and those that do not. The specific example of four-dimensional gravity coupled to a real massless scalar revealed a fascinating result which is otherwise quite robust. The solutions on such a ``critical surface'' develop indeed some form of space-time self-similarity or homothetic invariance, a phenomenon suggestively reminiscent of the emergence of conformal symmetry in ordinary critical statistical systems. This discovery has given rise to a wide research effort devoted to critical phenomena and scale invariance in gravity.

Choptuik's thought experiment~\cite{Chop} entails the ``tuning'' of initial conditions across the critical surface that separates solutions with a black-hole final state from those without it, focusing on the
pseudo-critical behavior attained by several observables when approaching this threshold~\cite{Chop,AlvarezGaume:2008fx}. Specifically, one considers a one-parameter class of initial conditions indexed by a value $p$ qualitatively related to the initial field amplitudes. For sufficiently small $p$, time evolution is linear and therefore no gravitational collapse occurs. For $p$ sufficiently large, a trapped surface forms, and therefore the final state must contain a black hole. Some critical value $p_\text{crit}$ should then signal the transition between these two regimes. Furthermore, a critical scaling law was exhibited; specifically, for $p>p_\text{crit}$ the mass of the resulting black hole scaled as

\begin{equation}
M_{\rm bh}(p) \propto (p-p_\text{crit})^\gamma\,,
\end{equation}

with the critical exponent (henceforth Choptuik exponent) $\gamma\approx0.37$ (see also~\cite{Gundlach,Hamade:1995ce}).

Numerical investigations showed that, close to criticality, the solutions exhibit discretely self-similar behaviors, which indicates that they are invariant under some finite homothety. Discrete self-similarity, while fascinating, is exceptionally difficult to study analytically. Non-trivial continuous self-similarity, i.e. the invariance under continuous one-parameter groups of homotheties, presents itself as soon as one introduces matter sectors with internal symmetries; it is then possible to obtain non-trivial solutions that are invariant under combinations of scalings and internal symmetry transformations.

One can perform similar numerical experiments with various choices for the matter content. For example,~\cite{HirschmannEardley} examined the case of a complex scalar field, while ~\cite{AlvarezGaume:2008qs,evanscoleman,KHA,MA} studied the critical collapse of a radiation fluid, and \cite{Hirschmann_1997} considered non-linear $\sigma$-models over the hyperbolic plane and the sphere. In all these cases, however, the focus was restricted to compact subgroups of the target-space symmetries. \cite{Hirschmann_1997} specifically overlaps with the contents of this paper, since it also examines elliptic solutions of the axion-dilaton system. The authors of~\cite{AE} also considered the generalization of the preceding results to the case of axial symmetries. All these results provide some evidence that the link between criticality of collapse and self-similarity extends well beyond the original context of spherically symmetric real scalar fields.

This paper is devoted to the study of continuously self-similar solutions of the Einstein-axion-dilaton system in arbitrary dimensions. The internal $\sltr$ symmetries have an interesting structure, due to the presence of compact and non-compact one-parameter subgroups, and we thus generalize previous discussions considering all possible classes of continuously self-similar collapses. In this fashion, we discover several families of critical spacetimes in various dimensions.

\section{Continuous self-similarity in the axion-dilaton system}

The system considered in this work is common to the low-energy effective action of type II string theory~\cite{SEN_1994,Schwarz_1995}. 
It consists of gravity coupled to a dilaton $\phi$ and an axion~$a$. 
The two scalars can be combined into a single complex field
$
\tau \equiv a + i e^{- \phi}$ spanning the upper half of the complex plane. 
The action in $d \geq 4$ dimensions is
\begin{equation}
S = \int d^d x \sqrt{-g} \left( R - \frac{1}{2} \frac{ \partial_a \tau
\partial^a \bar{\tau}}{(\Im\tau)^2} \right) \; .
\label{eaction}\end{equation}
where $R$ is the Ricci scalar\footnote{Here and in the following we use the ``mostly plus'' signature convention. We also remark that, in the standard convention, the kinetic term of $\tau$ does not include the factor of $\frac{1}{2}$.}, and the scalar sector is a non-linear sigma model over the hyperbolic plane $\mathbb{H}^2$ with reference to a half-plane chart. The equations of motion stemming from action~\eqref{eaction} are
\begin{equation}
\label{eq:efes}
R_{ab} = \tilde{T}_{ab} \equiv \frac{1}{4 (\Im\tau)^2} ( \partial_a \tau \partial_b
\bar{\tau} + \partial_a \bar{\tau} \partial_b \tau)\,,
\end{equation}

\begin{equation}\label{eq:taueom}
\nabla^a \nabla_a \tau +\frac{ i \nabla^a \tau \nabla_a \tau }{
\Im\tau} = 0 \,.
\end{equation}

The isometries of the target space form the group $\sltr$ of M\"obius transformations acting on $\tau$ as
\begin{equation}
\tau \rightarrow \frac{a\tau+b}{c\tau+d} \,,\quad a,b,c,d \in \mathbb{R}\,,\quad \det \mqty(a & b \\ c & d) = 1 \,.
\label{sltwo}\end{equation}
They are global internal symmetries of the action and of its equations of motion. In string theory this $\sltr$ symmetry
is reduced by quantum effects to a discrete $\mathrm{SL}(2,\mathbb{Z})$
subgroup, the general form of the S-duality conjectured to be a
non-perturbative symmetry of IIB strings~\cite{Font:1990gx,SEN_1994,Schwarz_1995}.

We are interested in examining some of the solutions at the critical threshold separating black hole end states from empty Minkowski ones. We note that if one sets either $a=0$ or $\phi=0$, up to a redefinition one is left with a single minimally-coupled scalar and thus to the original setup of~\cite{Chop}, which exhibits discrete self-similarity.

The continuous self-similarity is due to the presence of a
homothetic Killing vector $\xi$ that produces global scale
transformations so that
\begin{equation}
{\cal L}_{\xi} g_{ab} = 2 g_{ab} \; .
\label{metxi}\end{equation}
The issue is how $\tau$ must transform under $\xi$. The internal $\sltr$ symmetry of the scalar sector can be consistently combined with the spacetime homothety generated by $\xi = t \frac{\partial}{\partial t} + r \frac{\partial}{\partial r}$. All in all, $\tau$ must be homothety-invariant up to an $\sltr$ transformation:
\begin{equation}
{\cal L}_{\xi} \tau = \alpha_0 + \alpha_1 \tau + \alpha_2 \tau^2 \; ,
\label{tauxi}\end{equation}
with $\alpha_i \in \mathbb{R}$.

We shall henceforth confine our attention to spherically symmetric configurations. Generic $d$-dimensional spherically-symmetric spacetime metrics can be cast in the form:
\begin{equation}
	ds^2 = \left(1+u(t,r)\right)\left(- b(t,r)^2dt^2 + dr^2\right)
			+ r^2d\Omega_q^2 \; \,,
\label{eq:metricansatz}\end{equation}
with $q\equiv d-2$. Continuous self-similarity (\ref{metxi}) is then guaranteed if the two functions $u(t,r)$, $b(t,r)$ depend only on a scale invariant combination $z \equiv - \frac{r}{t}$:
\begin{equation}
	b(t,r) = b(z), \quad {\rm and} \quad u(t,r) = u(z)\,.
\label{zansatz}\end{equation}
We note that all such metrics, with the exception of the trivial Minkowski case, feature a genuine curvature singularity at the homotetic center $r=t=0$. 

One can now distinguish physically inequivalent cases referring to the conjugacy class of the infinitesimal M\"obius transformations that compensate the dilation as in eq.~\eqref{tauxi}. The trivial transformation can be excluded, since it implies a flat spacetime necessarily, as we shall see in section~\ref{sec:trivial_class}. One is therefore left with the three separate classes of elliptic, parabolic and hyperbolic solutions.

\subsection{Self-similar axion-dilaton fields}

\subsubsection{Elliptic class}

Up to an $SL(2,\mathbb{R})$ transformation, the general form of a solution of eq.~\eqref{tauxi} in the elliptic class is given by
\begin{equation}
 \tau(t,r)	=  i \frac{ 1 - (-t)^{i \omega} f(z) }{ 1 + (-t)^{i
\omega} f(z)} ,
\label{tauansatz_elliptic}
\end{equation}
with $f(z)$ a complex function satisfying $\abs{f(z)}<1$, and $\omega$ a real constant, which we can take to be positive\footnote{This is because the $\sltr$ map $\tau\rightarrow-\frac{1}{\tau}$ leaves the ansatz invariant up to $\omega\rightarrow-\omega$}. The compensating transformation is a rotation; it consequently manifests itself as a residual symmetry of $f(z)$,
\begin{equation}
    f(z) \rightarrow e^{i\theta}f(z)\,,
\end{equation}
which leaves the equations of motion invariant. In the language of equation \eqref{tauxi}, we have
\begin{equation}
    \mathcal{L}_\xi \tau = \frac{\omega}{2} \left( \tau^2 +1\right)\,.  
\end{equation}
One can justify the ansatz~\eqref{tauansatz_elliptic} noting that the simplest possible representative of the elliptic class in the upper-half plane obtains combining conformal mappings to and from the Poincar\'e disk with rotation in the latter, as follows:
\begin{equation}
    \tau(t,r) = i\frac{1-F(t,r)}{1+F(t,r)}\,,
\end{equation}
\begin{equation}
    F(t,r) = (-t)^{i\omega} f(z)\,.
\end{equation}

\subsubsection{Hyperbolic class}

The general solution of eq.~\eqref{tauxi} in the hyperbolic class is, up to a global $\sltr$ transformation,
\begin{equation}
\tau(t,r) = (-t)^{\omega} f(z) \,,
\label{tauansatz_hyperbolic}
\end{equation}
with $f(z)$ a complex function satisfying $\Im f(z)>0$, and $\omega$ a real constant which can be taken positive\footnote{Yet again, $\tau\rightarrow -\frac{1}{\tau}$ mantains the form of the ansatz but changes the sign of $\omega$.}. In this case the compensating transformation is a boost / hyperbolic translation, and correspondingly there is the residual symmetry
\begin{equation}
    f(z) \rightarrow e^\lambda f(z)\,,\quad \lambda \in \mathbb{R}\,,
\end{equation}
with infinitesimal form
\begin{equation}
    \mathcal{L}_\xi \tau = \omega \tau\,.
\end{equation}
We remark that~\cite{hatefialvarez1108,hatefialvarez1307} used a different parametrization

\begin{equation}
    \tau(t,r) = \frac{1-(-t)^w f(z)}{1+(-t)^w f(z)}
\end{equation}

which is actually related to~\eqref{tauansatz_hyperbolic} by an $\sltr$ transformation, and is thus entirely equivalent and results in the same equations of motion for $f(z)$.

\subsubsection{Parabolic class}

Finally, the general solution of eq.~\eqref{tauxi} in the parabolic class is of the form
 \begin{equation}
\tau(t,r) = f(z)+\omega \log(-t)\,,
\end{equation}
with $\Im f(z)>0$, and $\omega$ is real and can be taken positive for identical reasons as before. The infinitesimal form is
\begin{equation}
    \mathcal{L}_\xi \tau = \omega\,.
\end{equation}
From a geometric standpoint, this transformation is a horolation of $\mathbb{H}^2$; in practice it is a shift of $\tau$ by a real constant. This implies the following residual symmetry
\begin{equation}
    f(z) \rightarrow f(z) + a\,,
\end{equation}
under which the equations of motion will be invariant.

We also note a special type of transformation, unique to the parabolic case. Indeed, the transformation
\begin{equation}\label{parabolic_rescaling}
    \omega \rightarrow K \omega\,,\quad f(z) \rightarrow K f(z)\,,\quad K \in \mathbb{R}_+
\end{equation}
is equivalent to $\tau \rightarrow K \tau$, which is also a symmetry. Therefore, \eqref{parabolic_rescaling} will relate solutions with different $\omega>0$.

\section{Equations of motion}

We now turn to the reduced form of the equations of motion under the assumptions of continuous self-similarity and spherical symmetry.

\subsection{Methodology}

The determination of the final equations of motion for the given ans\"atze starting from the eqs.~\eqref{eq:efes},~\eqref{eq:taueom} in general dimension and for all three conjugacy classes is particularly arduous; thus, here we only sketch the strategy employed in the calculation.

\subsubsection{Dimensionally-agnostic calculation}

We consider the auxiliary two-dimensional metric:
\begin{equation}\label{eq:twodmetric}
    d\tilde{s}^2 = (1+u(t,r))\left(-b(t,r)^2 dt^2 + dr^2\right)
\end{equation}
for which we compute the two dimensional Ricci-tensor $\tilde{R}_{ab}$ and Christoffel symbols $\tilde{\Gamma}\indices{^a_b_c}$. This calculation needs to be performed only once. At this point, given that the metric~\eqref{eq:metricansatz} is written as a warped product of the $q$-sphere (we recall $q = d-2$)
\begin{equation}
    ds^2 = d\tilde{s}^2 + r^2 d\Omega_q^2\,,
\end{equation}
one can now easily derive relationships between the $d$-dimensional Ricci tensor and $\tilde{R}_{ab}$, $\tilde{\Gamma}\indices{^a_b_c}$ and the sphere's Ricci tensor, which is $R_{ij} = (q-1) h_{ij}$ where $h_{ij}$ is the unit sphere metric. Specifically:
\begin{eqaed}
    R_{tt} & = \tilde{R}_{tt} + q \frac{1}{r}\tilde{\Gamma}\indices{^r_t_t}\\
    R_{tr} & = \tilde{R}_{tr} + q \frac{1}{r}\tilde{\Gamma}\indices{^r_t_r}\\
    R_{rr} & = \tilde{R}_{rr} + q \frac{1}{r}\tilde{\Gamma}\indices{^r_r_r}\\
    R_{ij} & = \bigg((q-1) -r^2 \Big( \tilde{\Delta} \log r  + q \,\tilde{g}^{rr} \frac{1}{r^2} \Big)\bigg) h_{ij}\\
\end{eqaed}
where $\tilde{\Delta}$ is the two-dimensional Laplacian. All other components vanish.

The full Ricci tensor is hence determined explicitly as a function of the unknown dimension through $q$. In particular, we find
 \begin{eqaed}
    R_{tt} = \frac{1}{2 r b (u+1)^2}\,\bigg(& b^2 (u+1) \Big(\partial_r b(2 q u+r \partial_r u
    +2 q) \\& + 2 r \partial_r^2 b (u+1)\Big) +r \partial_t b (u+1) \partial_t u
   \\&  +b^3(q (u+1) \partial_r u-r \partial_r u^2+r (u+1) \partial_r^2 u) 
     \\& +r b(\partial_t u^2-(u+1) \partial_t^2 u\bigg)
 \end{eqaed}
 \begin{eqnarray}
    R_{tr} = \frac{q \partial_t u}{2 r (u+1)}
 \end{eqnarray}
\begin{eqaed}
R_{rr} = \frac{1}{2r b^3(1+u)^2} \bigg( & -rb^2 (1+u)\left(\partial_r b \partial_r u + 2 (1+u) \partial_r^2 b \right) \\ & + b^3 \big(q(1+u)\partial_r u + r (\partial_r u)^2 -r(1+u) \partial_r^2 u \big)\\ & - r(1+u) \,\partial_t b \,\partial_t u + rb \big(-(\partial_t u)^2 +(1+u) \partial_t^2 u \big) \bigg)
\end{eqaed}
\begin{equation}
    R_{ij} = \frac{(q-1) b u-r \partial_r b}{b (u+1)}\,h_{ij}\,,
\end{equation}
where $b$ and $u$ are the metric functions defined in eq.~\eqref{eq:twodmetric}.

The right hand side of the field equations
\begin{equation}
    \tilde{T}_{ab} = \frac{1}{4 \Im \tau^2} \left(\partial_a \tau \partial_b \bar{\tau} + \partial_a \bar \tau \partial_b \tau\right)\,, 
\end{equation}
it is instead readily seen to be dimension-invariant, so one needs to perform this calculation only once as well. The final piece is the axion-dilaton equation of motion, where the dimensional dependence appears in the $\nabla^a \nabla_a \tau$ term; if one however rewrites this expression as
\begin{equation}
    \nabla^a \nabla_a \tau = \frac{1}{\sqrt{-\det g}} \partial_a\left( \sqrt{-\det g} g^{ab} \partial_b \tau \right)\,,
\end{equation}
then this dependence only manifests itself in $\sqrt{-\det g} =r^q \sqrt{-\det \tilde{g}} $, so that, remembering the assumption of spherical symmetry, this equation too can be readily obtained as an explicit function of $q$.

Thus, all the equations of motion for $u$, $b$, $\tau$ are obtained for general dimension with a calculation of time complexity not scaling with the dimension itself. With this strategy, arbitrarily large dimensions could in principle be probed numerically with no increase in the machine time.

At this point, for the three conjugacy classes we can substitute the various ans\"atze for $\tau(t,r)$ in terms of $f(z)$.

\subsubsection{Elimination of \texorpdfstring{$u(z)$, $u'(z)$}{u(z), u'(z)}} \label{sec:elminateu}

In a self-similar solution, metric functions only depend on $z= -\frac{r}{t}$, so $u(t,r)=u(z)$ and $b(t,r) = b(z)$. One can then eliminate $u(z)$ in favour of $b(z)$ and $f(z)$. Specifically, the $ij$ Einstein field equations (EFEs) lead to
\begin{equation}\label{eq:u0explicit}
    u(z) = \frac{z b'(z)}{(q-1) b(z)}\,,
\end{equation}
while the first derivative is extracted from the $tr$ component of EFE. Specifically, in the elliptic, parabolic and hyperbolic cases 
\begin{equation}\label{eq:u0pexplicit}
    \frac{qu'(z)}{2(1+u(z))} = \begin{dcases}
    \frac{4zf'(z)\bar f'(z)+2i( w \bar f(z)f'(z) - w f(z)\bar f'(z))}{2(f(z)\bar f(z)-1)^2}  & \\
    \quad\quad\\
    \frac{w  (f'(z) + \bar f'(z)) - 2z  \bar f'(z) f'(z)}{(f(z)-\bar f(z))^2} & \\
     \quad\quad\\
    \frac{w \bar f(z) f'(z)+w  f(z) \bar f'(z) - 2z  \bar f'(z) f'(z)}{(f(z)-\bar f(z))^2} & 
    \end{dcases}
\end{equation}

One can then consider the combination of the field equations where second order derivative terms of the fields cancel, usually referred to as the Hamiltonian constraint
\begin{equation}
    R_{tt} + b^2\, R_{rr} = \tilde{T}_{tt} + b^2 \, \tilde{T}_{rr}\,,
\end{equation}
where $b$ is again the metric function defined in eq.~\eqref{eq:twodmetric}, replacing in it $u'$ and $u$ according to eqts.~\eqref{eq:u0explicit},~\eqref{eq:u0pexplicit}. After some algebra, this procedure yields a first-order differential equation for $b(z)$ involving $f(z)$ and $f'(z)$ (and conjugates).

One then moves to the axion-dilaton equation of motion~\eqref{eq:taueom}, again replacing $u(z)$ and $u'(z)$, and finally removing any instance of $b'(z)$ making use of the equation just found. The final result is a system of coupled ordinary differential equations of the form
\begin{align}
    b'(z) & = B(b(z),f(z),f'(z))\\
    f''(z) &= F(b(z),f(z),f'(z))
\end{align}
whose (real) order is five. We emphasize that we have been able to algebraically remove the degree of freedom associated with $u(z)$. We can now present the resulting system for the three conjugacy classes of $\sltra$.
 
\subsection{Final forms of the equations of motion}

Here we present the final equations of motion for $f(z)$, $b(z)$ for the three separate classes, for arbitrary numbers of dimensions. Our results are in complete agreement with~\cite{hatefialvarez1108}, where they were first derived.

\subsubsection{Elliptic class}

We find the following e.o.m.'s for self-similar solutions in the elliptic class in any dimension $d = q+2$:
\begin{eqnarray}
b' & = & \frac{ - { 2z(b^2 - z^2)} f' \bar{f}' + {
2i \omega (b^2 - z^2) } (f \bar{f}' - \bar{f} f')
+ {2\omega^2 z |f|^2 }}{q b (1-\abs{f}^2)^2} \,, \nonumber\\
\end{eqnarray}
%
%
%
\begin{eqaed}
q z  \left(z^2-b^2\right) (1-\abs{f}^2)^2 f'' = & \,\, b^2 f'  \big( -2 f \left(q z \bar{f}^2 f'-i \omega z \bar{f}'+q^2 \bar{f}\right) \\ 
& \quad -2 z^2 f' \bar{f}'+2 z \bar{f}  (q-i \omega) f'+q^2 \abs{f}^4+q^2\big)\\
& +z  \Big(2 f^2 \left(q (-1-i \omega) z \bar{f}^2 f' +\omega^2 z \bar{f}' -i q \omega \bar{f} \right)\\
&  \quad + f  \left(2 i \omega z^2 f'  \bar{f}' +2 q z^2 \bar{f}^2 f'^2+4 q z \bar{f}  f' +q \omega (\omega+i)\right)\\
&\quad -2 q z f'  \left(z \bar{f}  f' -i \omega+1\right)-q \omega (\omega-i) \abs{f}^2 f\Big)\\
& +\frac{2 z^3}{b^2} \left(z f' -i \omega f \right)^2 \left(z \bar{f}' +i \omega \bar{f} \right)
\end{eqaed}

\subsubsection{Parabolic class}

We find the following e.o.m.s for self-similar solutions in the parabolic class, in any dimension $d = q+2$:
%
\begin{eqaed}
    b' = -\frac{2 \left(\left(z^2-b^2\right) f' \left(z \bar{f}'-\omega\right)+\omega \left(\left(b^2-z^2\right) \bar{f}'+\omega z\right)\right)}{q b (f-\bar{f})^2}
\end{eqaed}
\begin{eqaed}
    {q z\left(z^2-b^2\right) (f-\bar{f})^2} f''  = &\,\, b^2 f' \big(2 z f' \left(z \bar{f}'-\omega\right)-2 q f \left(z f'+q \bar{f}\right)\\
    &\quad + 2 q z \bar{f} f'+q^2 f^2-2 \omega z \bar{f}'+q^2 \bar{f}^2\big)\\
    & +z  \Big(2 \omega z \bar{f}' \left(\omega-z f'\right)+2 q f \big(\left(\omega-z f'\right)^2-\bar{f} \left(\omega-2 z f'\right)\big)\\
    &\quad -2 q \bar{f} \left(\omega-z f'\right)^2  +q \bar{f}^2 \left(\omega-2 z f'\right)+q f^2 \left(\omega-2 z f'\right)\Big)\\
    & +\frac{2 z^3}{b^2} \left(\omega-z f'\right)^2 \left(\omega-z \bar{f}'\right)
\end{eqaed}

\subsubsection{Hyperbolic class}

Finally, we find the following e.o.m.s for self-similar solutions in the hyperbolic class, in any dimension $d = q+2$:
%
%
\begin{eqaed}
    b' = -\frac{2 \left(\left(z^2-b^2\right) f' \left(z \bar{f}'-\omega \bar{f}\right)+\omega f \left(\left(b^2-z^2\right) \bar{f}'+\omega z \bar{f}\right)\right)}{q b (f-\bar{f})^2}
\end{eqaed}
\begin{eqaed}
   {q z  \left(z^2-b^2\right) (f-\bar{f})^2} f'' = &\,\, b^2 f' \big(2 z^2 f'\bar{f}'-2 z \omega f \bar{f}'-2\, q\, f (z f'+q \bar{f})\\
   & \quad +2\, q \,z \bar{f} f'+q^2 f^2-2 \omega z \bar{f}{f}'+q^2 \bar{f}^2\big)\\
   +&z \Big(q\omega (1+\omega)f^3-2qz\bar ff'(\bar f-\omega\bar f+zf')\\
   &-2f^2\big(q\omega\bar f+q(1+\omega)zf'-\omega^2z\bar f'\big)\\&
   +f(-q(-1+\omega)\omega\bar f^2+4qz\bar ff'+2z^2f'(qf'-\omega\bar f'))\Big)\\
   &+\frac{2 z^3}{b^2} \left(\omega f-z f'\right)^2(\omega \bar{f}-z \bar{f}')
\end{eqaed}

\section{Self-similar solutions}

Having reduced the problem of determining continuously self-similar (CSS) configurations to a one-dimensional system of ODEs for the functions $b(z)$, $f(z)$, we now turn to the question of searching physically meaningful solutions to the latter.

\subsection{Search methodology}\label{sec:search_method}

In~\cite{Hirschmann_1997} the authors present an accurate numerical method for the search of self-similar solutions. We build our technique upon theirs, with several important simplifications on the algebraic and numerical fronts.

We begin from the endpoint $z=0$, which is a singular point for the system. In terms of the original $t$, $r$ coordinates, this is the ray $r=0$, $t<0$. Therefore, it must be a mere coordinate singularity, specifically of the type associated with the origin of a polar coordinate system. If we thus assume that the scalars are regular across this axis, it follows that
\begin{equation}
    f'(0) =0\,,
\end{equation}
which comprises two real boundary conditions.

In addition, making use of the freedom to rescale time by an overall constant, one can set
\begin{equation}
    b(0) = 1\,.
\end{equation}
Finally, we recall that in each conjugacy class the equations of motion have a residual one-dimensional symmetry, which we employ to remove one degree of freedom out of the complex number $f(0)$. Specifically, we perform the following fixing:
\begin{equation}\label{eq:zeroffixing}
    f(0) = \begin{cases}
        \abs{f(0)} \in \mathbb{R}_+ & \mathrm{elliptic}\\
        \quad\quad\\
        i \Im f(0)\,,\quad \Im f(0) \in \mathbb{R}_+ & \mathrm{parabolic}\\
         \quad\quad\\
        1 + i \Im f(0)\,,\quad \Im f(0) \in \mathbb{R}_+ & \mathrm{hyperbolic}
    \end{cases}
\end{equation}
We also considered the equivalent fixing $f(0) = e^{i \arg f(0)}$ for the hyperbolic class, but we ultimately preferred the choice presented in \eqref{eq:zeroffixing} for consistency with the existing literature. At any rate, the problem is reduced to the determination of two real parameters, the remaining one in $f(0)$ and $\omega$.

Note that another kind of singularity is developed whenever the homothetic Killing vector $\xi$ is null, namely when $b^2(z) - z^2$ vanishes. In fact, this does first happen for all non-trivial initial conditions at some $z_+$ where
\begin{equation}
    b(z_+) = z_+\,.
\end{equation}
We call the hypersurface $z=z_+$, which is itself null, a homothetic horizon. Since it is just a past horizon for the genuine singularity at the homotetic center, it cannot be physical in any sense, and it must also be a mere coordinate singularity. We therefore demand that the axion-dilaton be regular across it. This is equivalent, in fact, to expanding the equations of motion near $z_+$, which is a singular point for the $f(z)$ equation, and imposing that $f''(z)$ stay finite as $z\rightarrow z_+$. The cancellation of the divergent part of $f''(z)$ is a complex-valued constraint $G$ involving the values of the functions at $z_+$:
\begin{equation}\label{generic_constraint_zplus}
    \mathbb{C}\, \ni \,G(f(z_+), f'(z_+), b(z_+)) = 0\,,
\end{equation}
whose specific form we shall present shortly, for each conjugacy class, in section~\ref{sec:explicitG}. With an equal number of real constraints and real unknowns, the system is completely determined, and the solution set is generically discrete.

The aforementioned correspondence between parameters and constraints fails for the parabolic class, because of the exceptional symmetry described in~\eqref{parabolic_rescaling}. In this case, if the parameters $(\omega,\Im f(0))$ generate a solution, so do $(K \omega,K\Im f(0))$, since both the equations of motion and the condition \eqref{generic_constraint_zplus} are invariant under this rescaling. Because of this degeneracy, the only real physical unknown parameter is therefore the ratio $\omega/\Im f(0)$, and the two-dimensional constraint must be solved for it. The system is thus a priori over-constrained, and part of the procedure must be reconsidered, as we shall see.

In practice, our numerical procedure is as follows. Let $x_0$ be the undetermined real parameter in $f(0)$, which can be $\abs{f(0)}$ or $\Im f(0)$ depending on the class. We define a complex-valued ``mismatch function'' of $\omega$ and $x_0$ as follows:

\begin{itemize}
    \item Determine $f(0)$ from $x_0$ for the specific class of interest according to~\eqref{eq:zeroffixing}.
    \item Having complete boundary conditions at $z=0$, integrate the equations of motion forward numerically, starting from a small positive $z_0$ to avoid the singularity at $z=0$.
    \item Stop the integration whenever the function $b(z)-z$ is first detected to dip below a threshold value $\delta$, which is positive but small. The precise $z$ where the crossing happens is identified as the corresponding estimate for $z_+$.
    \item Using the numerical solution, evaluate $f(z_+)$, $f'(z_+)$, $b(z_+)$, and output the constraint $G$ as defined in \eqref{generic_constraint_zplus}.
\end{itemize} 

Up to the inaccuracies introduced by the regularization parameters $z_0$ and $\delta$, the above $G(\omega,x_0)$ function will then vanish if and only if the resulting solution is regular at the homothetic horizon. Therefore, the problem of identifying regular solutions is translated into a search for zeroes of a complex function of two real variables. In practice, we plot curves where $\Re G$ and $\Im G$ vanish in the $(\omega,x_0)$ plane, and look for their intersections visually. After a suitable starting guess is thus identified, standard root-finding procedures can provide much more precise estimates for the positions of the roots.

Since the output of the root-finding procedure is especially accurate (in most cases we have verified that $\abs{G}$ can be reduced to at least $\sim 10^{-13}$), the major source of inaccuracy in the solution parameters can be ascribed to the magnitude of $z_0$ and $\delta$.

In the parabolic case, the aforementioned degeneracy reflects itself in the fact that the function $G(\omega,\Im f(0))$ only depends on the ratio $\omega/\Im f(0)$. Two-dimensional root-finding would be unsuitable in this case, which always features a flat direction, and fails to determine solutions correctly. One can instead trace the zeroes of, say, $G(\omega,1)$ for varying $\omega$. Any such root $\omega^*$, if it exists, would then generate a continuous ray of solutions $(\omega,\Im f(0)) = (K \omega^*, K)$. Different solutions within the same ray are related by an $\sltr$ rescaling, and are thus physically indistinguishable.

\subsubsection{Global structure of spacetime}

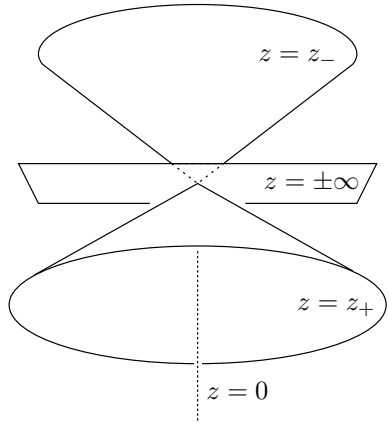
\begin{figure}
    \centering
\begingroup%
  \makeatletter%
  \providecommand\color[2][]{%
    \errmessage{(Inkscape) Color is used for the text in Inkscape, but the package 'color.sty' is not loaded}%
    \renewcommand\color[2][]{}%
  }%
  \providecommand\transparent[1]{%
    \errmessage{(Inkscape) Transparency is used (non-zero) for the text in Inkscape, but the package 'transparent.sty' is not loaded}%
    \renewcommand\transparent[1]{}%
  }%
  \providecommand\rotatebox[2]{#2}%
  \newcommand*\fsize{\dimexpr\f@size pt\relax}%
  \newcommand*\lineheight[1]{\fontsize{\fsize}{#1\fsize}\selectfont}%
  \ifx\svgwidth\undefined%
    \setlength{\unitlength}{151.09053724bp}%
    \ifx\svgscale\undefined%
      \relax%
    \else%
      \setlength{\unitlength}{\unitlength * \real{\svgscale}}%
    \fi%
  \else%
    \setlength{\unitlength}{\svgwidth}%
  \fi%
  \global\let\svgwidth\undefined%
  \global\let\svgscale\undefined%
  \makeatother%
  \begin{picture}(1,1.04391204)%
    \lineheight{1}%
    \setlength\tabcolsep{0pt}%
    \put(0,0){\includegraphics[width=\unitlength,page=1]{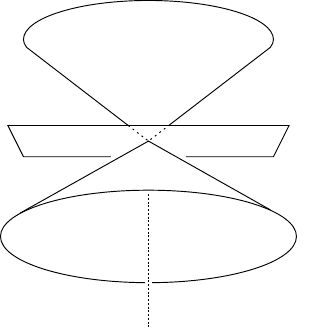}}%
    \put(0.392226,0.11913389){\color[rgb]{0,0,0}\makebox(0,0)[lt]{\begin{minipage}{0.33754595\unitlength}\raggedright \end{minipage}}}%
    \put(0.57117566,0.05894022){\color[rgb]{0,0,0}\makebox(0,0)[t]{\lineheight{1.25}\smash{\begin{tabular}[t]{c}$z=0$\end{tabular}}}}%
    \put(0.81759838,0.27965702){\color[rgb]{0,0,0}\makebox(0,0)[t]{\lineheight{1.25}\smash{\begin{tabular}[t]{c}$z=z_+$\end{tabular}}}}%
    \put(0.75643591,0.58192378){\color[rgb]{0,0,0}\makebox(0,0)[t]{\lineheight{1.25}\smash{\begin{tabular}[t]{c}$z=\pm\infty$\end{tabular}}}}%
    \put(0.72186578,0.90103229){\color[rgb]{0,0,0}\makebox(0,0)[t]{\lineheight{1.25}\smash{\begin{tabular}[t]{c}$z=z_-$\end{tabular}}}}%
  \end{picture}%
\endgroup%

    \caption{Schematic structure of the geometry of a non-trivial self-similar metric, with singular loci of the $(z,t)$ chart mapped.}
    \label{fig:geometry}
\end{figure}

We follow the analysis in~\cite{hatefialvarez1108,hatefialvarez1307}. We remark that $z=0$ and $z=z_+$ are not the only singularities of the equations of motion or equivalently of the $(z,t)$ coordinate chart. We have, in total, five singularities, as depicted in Figure~\ref{fig:geometry}:

\begin{eqaed}
    z & =  0\\
    z & =  z_+>0\,,\quad b(z_+) = z_+\\
    z & =  \pm \infty \\
    z & =  z_-<0\,,\quad b(z_-) = - z_-
\end{eqaed}

We have already discussed $0$ and $z_+$. The surface $z=\pm\infty$, or $t=0$, is clearly also a coordinate singularity, and it is removed demanding that the functions propagate regularly across it from $z\rightarrow +\infty$ to $z\rightarrow -\infty$. Since this surface is not null, this requirement does not involve any additional constraints. Therefore solutions for $0<z<z_+$ can be smoothly continued up to this surface through $z_+ < z < +\infty$, and then across it for $-\infty < z$.

The last singularity is encountered when $b(z_-) = - z_-$, which happens at some $z_-<0$. This is the future light cone of the homotetic center, and like the first homothetic horizon it is null; therefore regularity across it imposes additional constraints which cannot generically be solved. Therefore, continuing solutions beyond $z_-$ is not possible, and thus it must be the location of an actual singularity bounding the region of existence and uniqueness for the equations of motion, a Cauchy horizon. In other words, note that points after $z_-$ can receive signals from the singularity at the homothetic center, which reflects a loss of uniqueness for the initial-value problem.

It is therefore reasonable to cut off solutions at the Cauchy horizon. In conclusion, a solution is valid in the range $0<z<z_+$ if and only if it has a valid extension from $z=0$ to $z=z_-$, which is as global a range as possible. This fact justifies our focus on the $0<z<z_+$ region only.

\subsubsection{Constraint at \texorpdfstring{$z_+$}{z+}}\label{sec:explicitG}

In this section we present the explicit form of $G(f(z_+),f'(z_+),b(z_+))$, which is proportional, in all cases, to the divergent part of $f''(z)$ as $z\rightarrow z_+$, and whose vanishing is equivalent to the regularity condition for the axion-dilaton through the horizon. The overall normalization of this function is of course arbitrary.

For the elliptic class:
\begin{eqaed}\label{eq:Gelliptic}
     G(f(z_+),f'(z_+)) = & \, 2 z \bar{f}(z_+) \left(q^2-2 q-2 \omega^2\right) f'(z_+)\\ & +f(z_+) \bar{f}(z_+) \left(q z_+ \bar{f}(z_+) (-q+2 i \omega+2) f'(z_+)+2 i \omega \left(q+\omega^2\right)\right)\\&-\frac{q z_+ (q+2 i \omega-2) f'(z_+)}{f(z_+)}\\&+q \omega (\omega-i) f(z_+)^2 \bar{f}(z_+)^2-q \omega (\omega+i)\,.
\end{eqaed}
%
%
For the parabolic case:
\begin{eqaed}\label{eq:Gparabolic}
G(f(z_+),f'(z_+))   = & \, -\frac{1}{f(z_+)}\Bigg(-2 q \omega \bar{f}(z_+) \left(\omega-2 z_+ f'(z_+)\right)\\& +q \bar{f}(z_+)^2 \left((q-2) z_+ f'(z_+)+\omega\right)\\&-2 q f(z_+) \left(\bar{f}(z_+) \left((q-2) z_+ f'(z_+)+\omega\right)-\omega \left(\omega-2 z_+ f'(z_+)\right)\right)\\&+q f(z_+)^2 \left((q-2) z_+ f'(z_+)+\omega\right)+2 \omega^2 \left(\omega-2 z_+ f'(z_+)\right)\Bigg)\,.
\end{eqaed}
Note that~\eqref{eq:Gparabolic} scales homogeneously as $K^2$ under $(\omega,f)\rightarrow (K\omega,Kf)$, so that its zeroes are invariant.
For the hyperbolic case:
\begin{eqaed}
    G(f(z_+),f'(z_+)) = &\, \bar{f}(z_+) \left(2 z_+ \left(q^2-2 q+2 \omega^2\right) f'(z_+)+q (\omega-1) \omega \bar{f}(z_+)\right)\\& +f(z_+) \left(q z_+ (-q+2 \omega+2) f'(z_+)+2 \omega \bar{f}(z_+) \left(q-\omega^2\right)\right)\\& -\frac{q z_+ \bar{f}(z_+)^2 (q+2 \omega-2) f'(z_+)}{f(z_+)}-q\, \omega (\omega+1) f(z_+)^2\,.
\end{eqaed}

\subsubsection{Trivial class}\label{sec:trivial_class}

It is also instructive to comment on the case of a truly self-similar axion-dilaton field $\tau$, that is to say where the compensating transformation of~\eqref{tauxi} vanishes. In other words, $\tau(t,r) = \tau(z)$. This case can be obtained from any of the previous ones sending $\omega \rightarrow 0$. Note that by identical arguments as before $b(0)=1$ $\tau'(0) = 0$, and $\tau(0) = \tau_0$ is an arbitrary constant, which can be set to any value in the upper-half complex plane by an $\sltr$ transformation. These are, however, the initial conditions that generate a flat spacetime with a constant axion-dilaton $\tau_0$. Therefore one can argue, by uniqueness of the solutions to the system of ODEs for this trivial class, that all the corresponding solutions are trivial.

\subsection{Results}

As an illustration, we can display explicit solutions in four and five dimensions for all three classes, which were obtained with the techniques described so far. Now we present our results for the corresponding six cases.

\subsubsection{Solutions for \texorpdfstring{$d=4$}{d=4}, elliptic class}

\begin{figure}[H]
    \centering
    \includegraphics[width=3in]{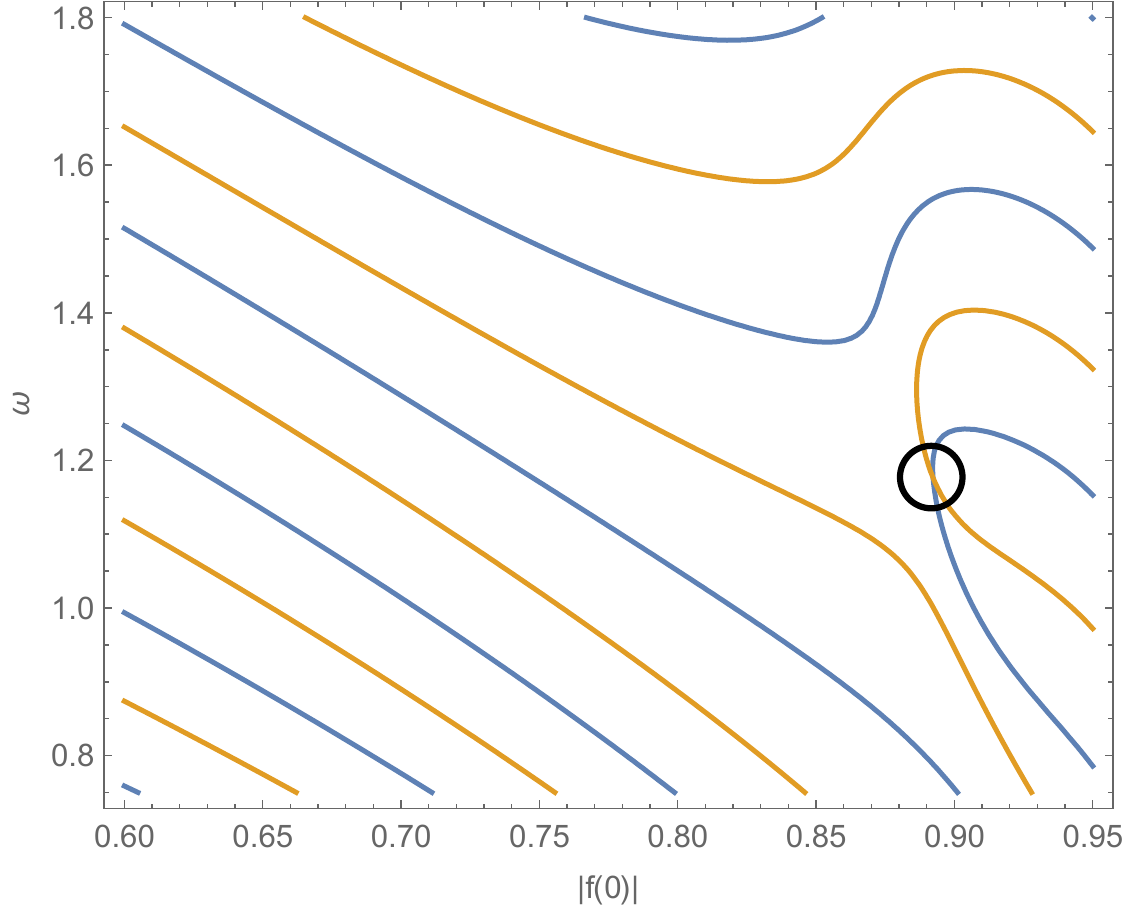}
    \caption{Curves of vanishing real and imaginary parts of $G(\omega,\abs{f(0)})$ in the $d=4$ elliptic case. Only one solution can be spotted as an intersection. No other solutions have been found outside the range of this plot.}
    \label{fig:4delliptic}
\end{figure}

\begin{figure}
    \centering
    \includegraphics[width=3in]{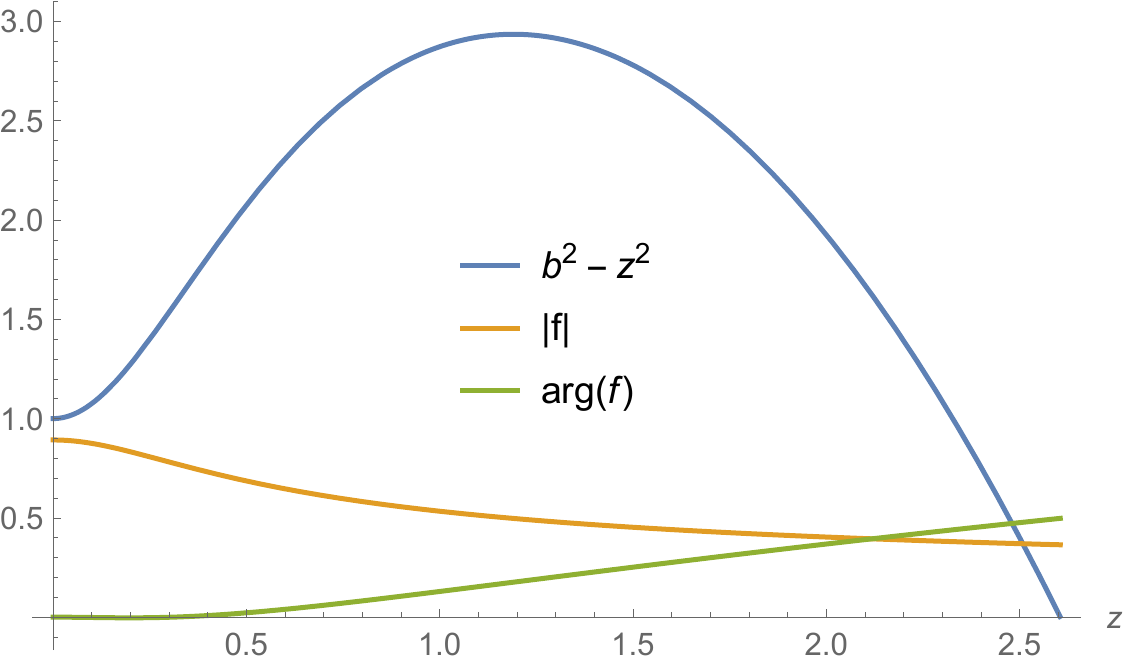}
    \caption{Profile of the single elliptic solution in four dimensions. We also plot the useful quantity $b^2(z)-z^2$, which is the redshift factor $g_{zz}$ in the $(z,t)$ coordinate chart and vanishes at the homothetic horizon.}
    \label{fig:sol4delliptic}
\end{figure}

Using the diagram in Figure~\ref{fig:4delliptic}, we are able to identify a single solution in the $d=4$ elliptic class, whose parameters are as follows:
\begin{eqaed}
    w & = 1.176 \\ \abs{f(0)} &= 0.892 \\ z_+ &= 2.605
\end{eqaed}
and whose profile is represented in Figure~\ref{fig:sol4delliptic}.

This solution was already known in the literature, see~\cite{Hirschmann_1995,Hirschmann_1997,hatefialvarez1108}.

\subsubsection{Solutions for \texorpdfstring{$d=4$}{d=4}, parabolic class}

\begin{figure}[H]
    \centering
    \includegraphics[width=2.5in]{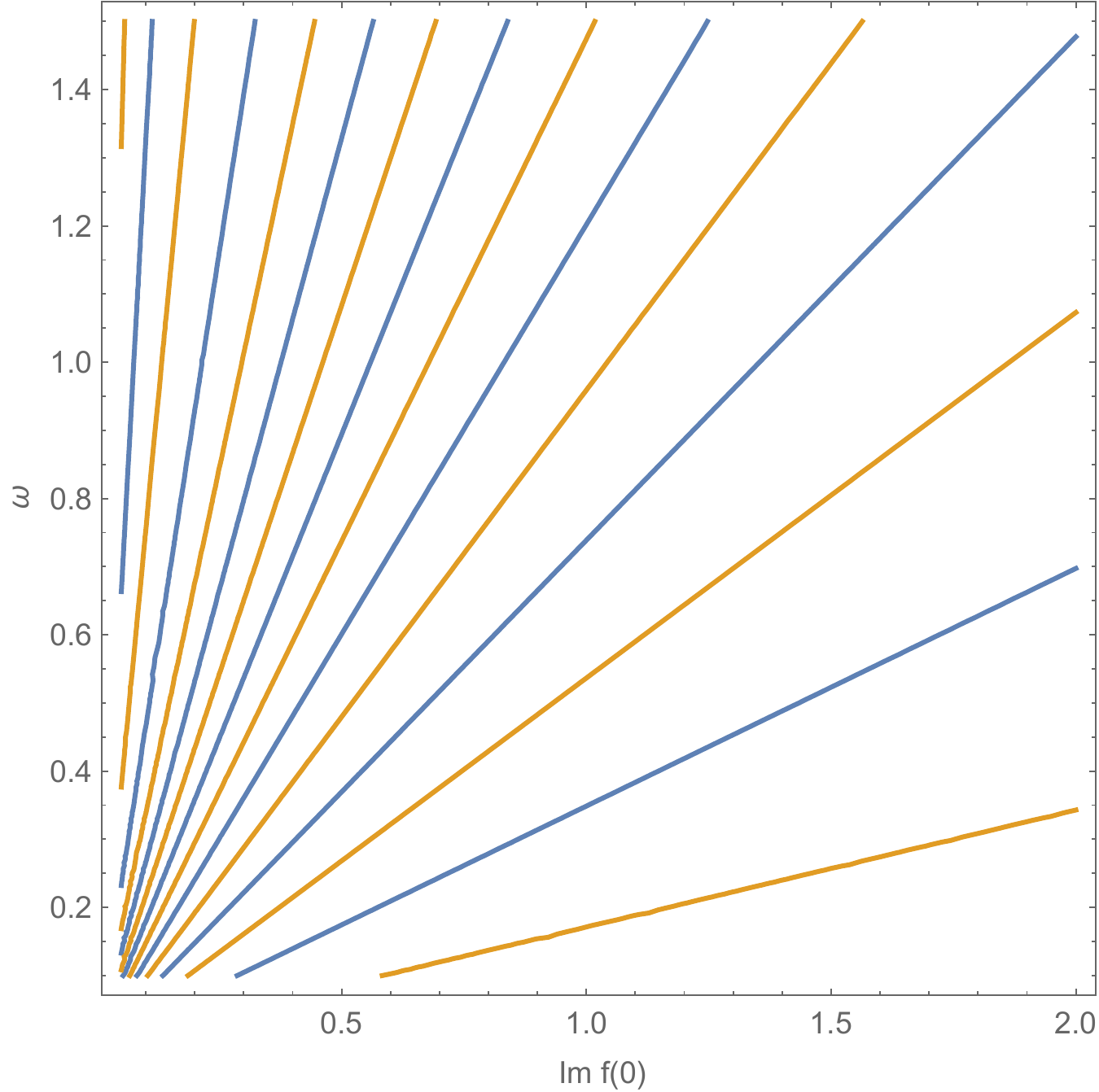}
    \caption{A two-dimensional plot of the zeroes of the real and imaginary parts of $G(\omega,\Im f(0))$, which highlights the degeneracy associated with the scaling invariance of the parabolic case. It is not possible to identify solution rays from this type of diagram.}
    \label{fig:testparabolic}
\end{figure}

\begin{figure}[H]
    \centering
    \includegraphics[width=3in]{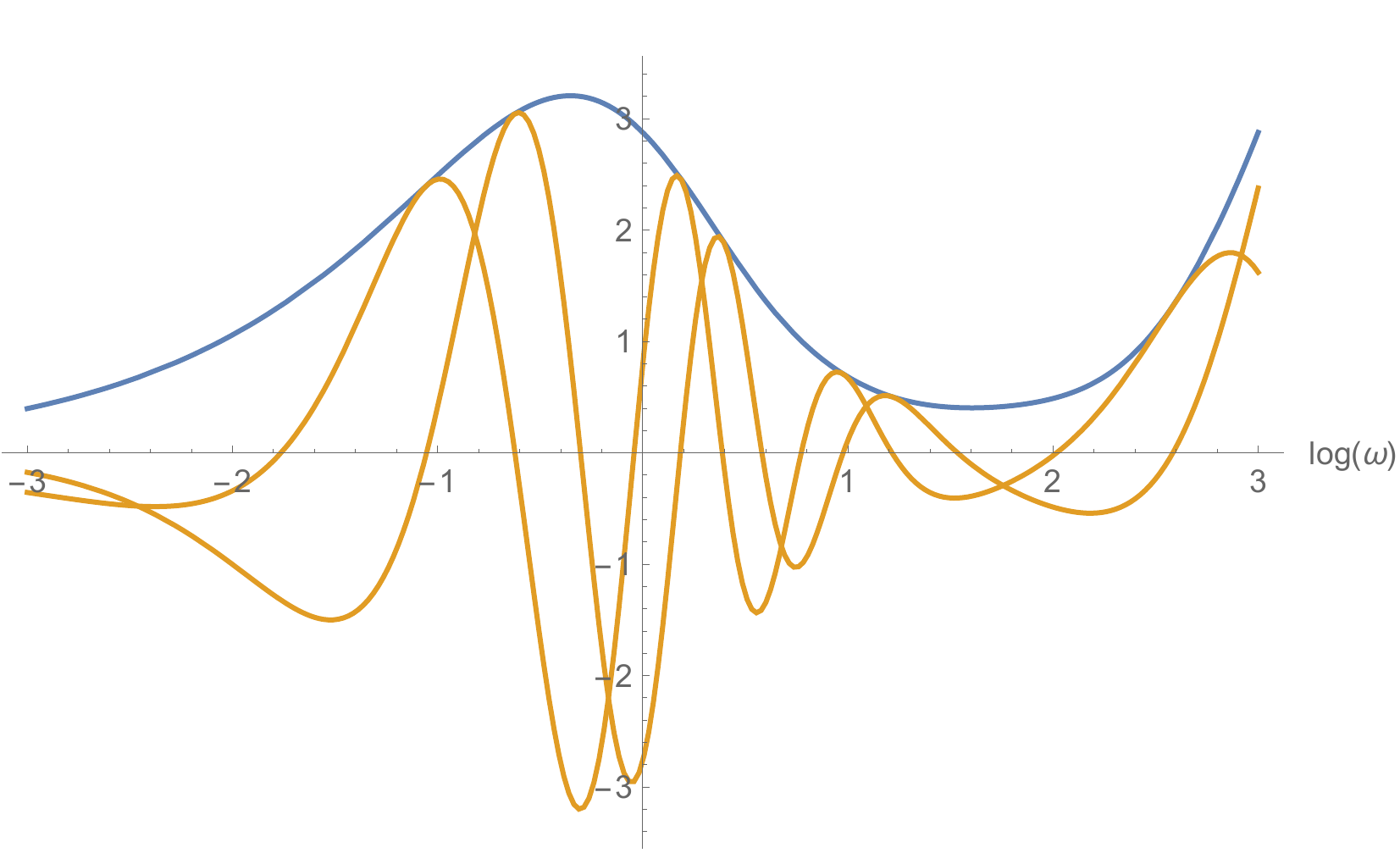}
    \caption{Plots of absolute value (blue) and real and imaginary parts (orange) of $G(\omega,1)$ over $\omega$ in the $d=4$ parabolic case. No zero is observed for $\omega > 0$.}    \label{fig:4dparabolic}
\end{figure}
Consistently with the arguments presented in section~\ref{sec:search_method} on the scaling symmetry, and reflected in Figure~\ref{fig:testparabolic}, in the parabolic class one just needs to find zeroes of $G(\omega,\Im f(0))$ over the single real parameter $\omega/\Im f(0)$. In Figure~\ref{fig:4dparabolic} we plot this complex function over $\omega$ for $\Im f(0)=1$. We have not been able to identify any roots in this case.

\subsubsection{Solutions for \texorpdfstring{$d=4$}{d=4}, hyperbolic class}

\begin{figure}
    \centering
    \includegraphics[width=\textwidth]{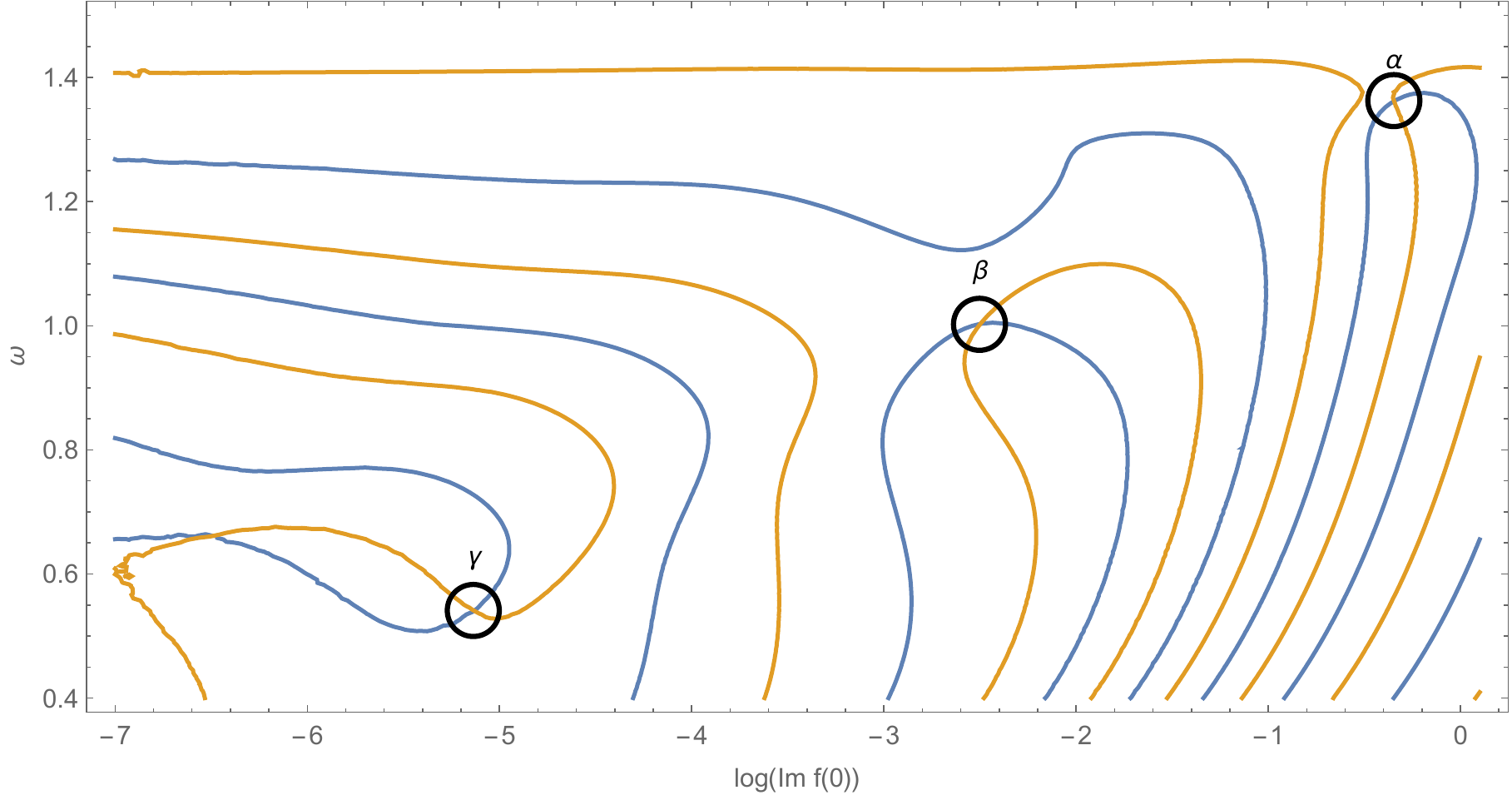}
    \caption{Curves of vanishing real and imaginary parts of $G(\omega,\Im f(0))$ in the $d=4$ hyperbolic case. Three solutions are easily spotted, which we call $\alpha$, $\beta$, $\gamma$ in order of decreasing $\Im f(0)$. A fourth intersection $\delta$ might lie around $\log \Im f(0) \sim -6.5$, $\omega \sim 0.65$, but we don't have enough control on the numerics to be sure.}
    \label{fig:4hyperbolic}
\end{figure}
We are able to identify several solutions of very rapidly decreasing $\Im f(0)$ in this class. No solutions are found with $\Im f(0)>1$. Moving to lower values, the first two solutions $\alpha$ and $\beta$ are clearly determined; for the third $\gamma$ solution we remark that due to the very small size of $\Im f(0)$ and correspondingly growing $z_+$ root-finding is affected by modest numerical noise and the quality is not excellent ($G \sim 10^{-7}$ as compared to the more typical $G \sim 10^{-13}-10^{-17}$), but we are confident in its existence. We tentatively also guess that a fourth root $\delta$ exists at a very small $\log \Im f(0) \sim -6.5$, $\omega \sim 0.65$, but numerical errors are already too large to be certain. We cannot investigate any further down with the current setup, and therefore we are unable at this point to make a definite statement as to whether there are any missing solutions.
\begin{figure}
    \centering
    \includegraphics[width=3.5in]{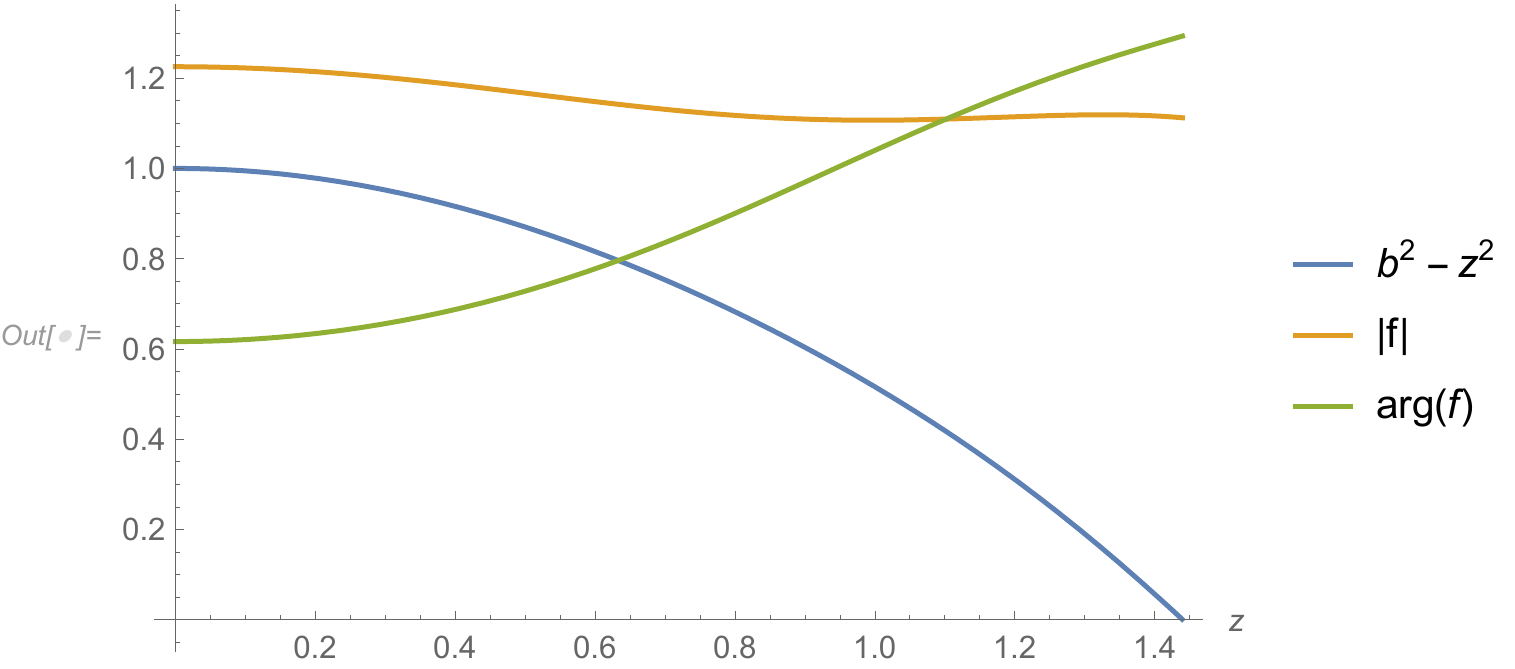}
    \caption{Profile of the $\alpha$ solution of the hyperbolic class in four dimensions}
    \label{fig:solution4dhyperbolicalpha}
\end{figure}
The solution parameters are reported as follows, with the caveat that the shaded row corresponds to a solution that is not verified to comforting accuracy. As an example, in Figure~\ref{fig:solution4dhyperbolicalpha} we display the profile of the $\alpha$ solution.
\begin{center}
\bgroup
\def\arraystretch{2}

\begin{tabular}{|c|c|c|c|}\hline
    Solution & $w$ & $\Im f(0)$ & $z_+$ \\ \hline
     $\alpha$ & $1.362$ & $0.708$ & $1.440$\\
     $\beta$ & $1.003$ & $0.0822$ & $3.29$ \\
     \rowcolor{light-gray}
     $\gamma$ & $0.541$ & $0.0059$ & $8.44$\\
     \hline
\end{tabular}
\egroup
\end{center}
\subsubsection{Solutions for \texorpdfstring{$d=5$}{d=5}, elliptic class}

\begin{figure}[H]
    \centering
    \includegraphics[width=2in]{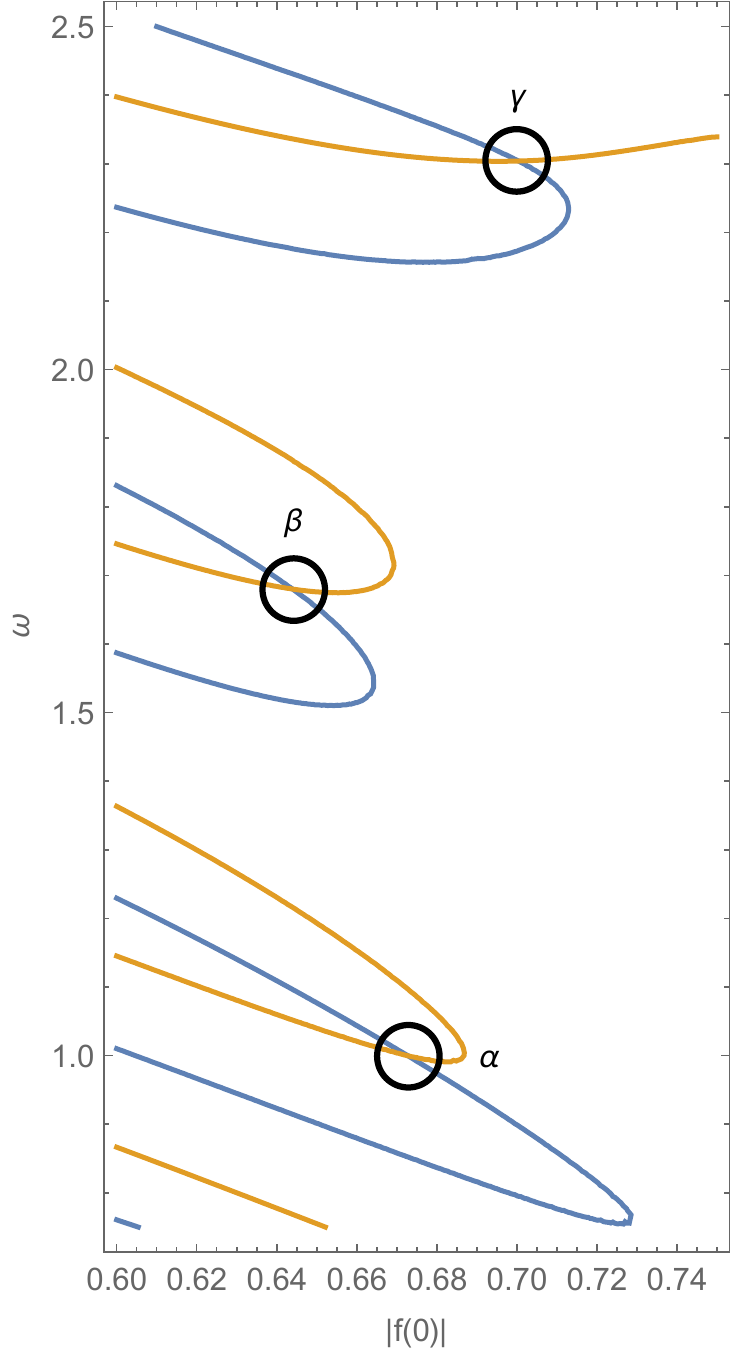}
    \caption{Curves of vanishing real and imaginary parts of $G(\omega,\abs{f(0)})$ in the $d=5$ elliptic case. Three solutions were spotted, which we call $\alpha$, $\beta$, $\gamma$ in order of increasing $\omega$. No other solutions were found outside this range.}
    \label{fig:5delliptic}
\end{figure}

In this case we have identified three distinct solutions, as displayed in Figure~\ref{fig:5delliptic}:

\begin{center}
\bgroup
\def\arraystretch{2}

\begin{tabular}{|c|c|c|c|}\hline
      Solution & $w$ & $\abs{f(0)}$ & $z_+$ \\
    \hline
    $\alpha$ & $0.999$ & $0.673$ & $1.246$\\
    $\beta$ & $1.680$ & $0.644$ & $1.397$ \\
    $\gamma$ & $2.304$ & $0.700$ & $1.694$\\
    \hline
\end{tabular}
\egroup
\end{center}

\subsubsection{Solutions for \texorpdfstring{$d=5$}{d=5}, parabolic class}

\begin{figure}[H]
    \centering
    \includegraphics[width=4in]{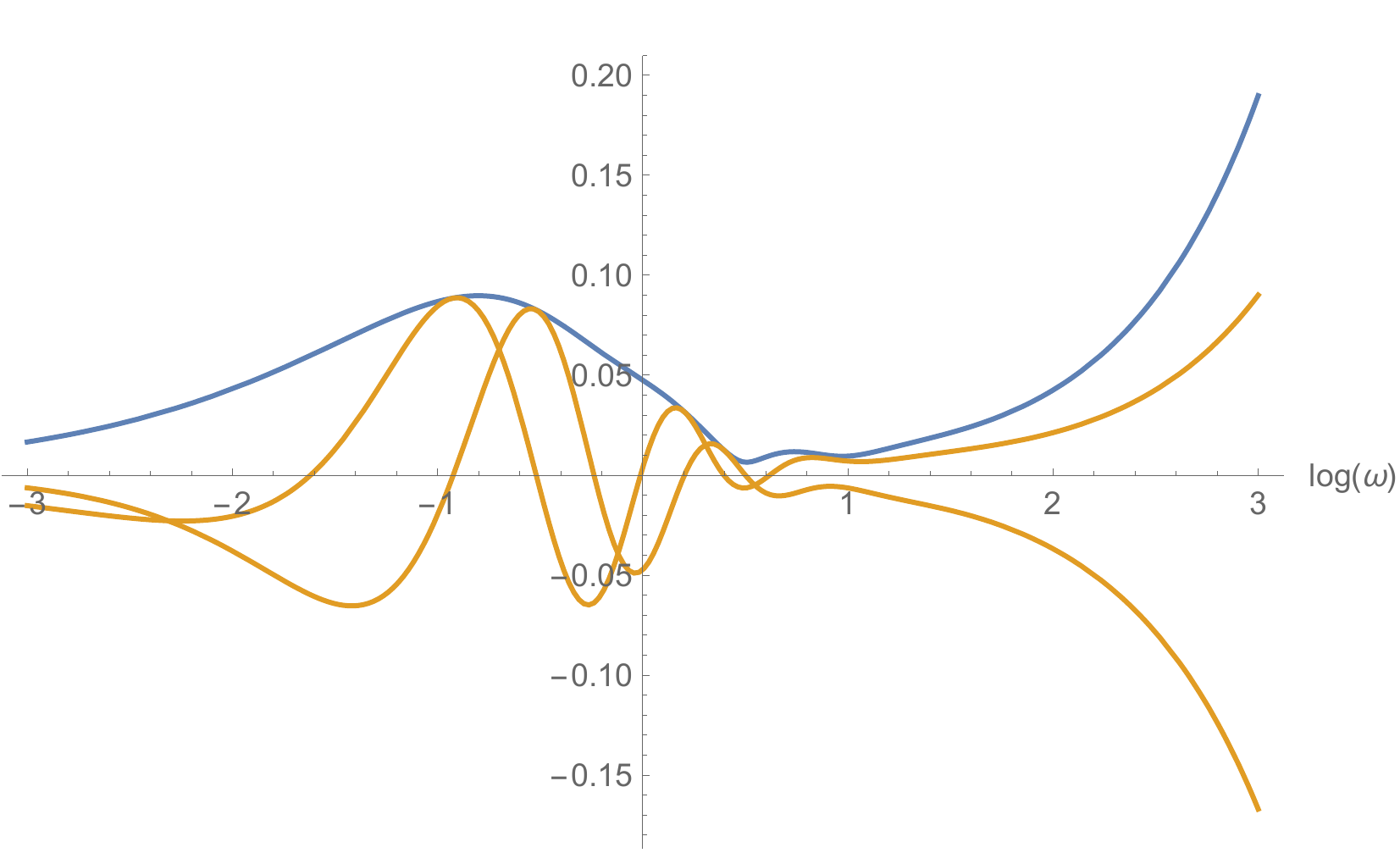}
    \caption{Plot of absolute value (blue), real and imaginary parts (orange) of $G(\omega,1)$ in the $d=5$ parabolic case.}
    \label{fig:my_label}
\end{figure}

We do not observe any solutions in this class, as in four dimensions. We do however note the remarkably low value of $\abs{G} \sim 0.006$ around $\omega \sim 1.65$. At this moment, we do not exclude that a solution ray in five dimensions that might have been rendered invisible by numerical errors, which is a possible option for such an over-determined system.

\subsubsection{Solutions for \texorpdfstring{$d=5$}{d=5}, hyperbolic class}

\begin{figure}[H]
    \centering
    \includegraphics[width=4in]{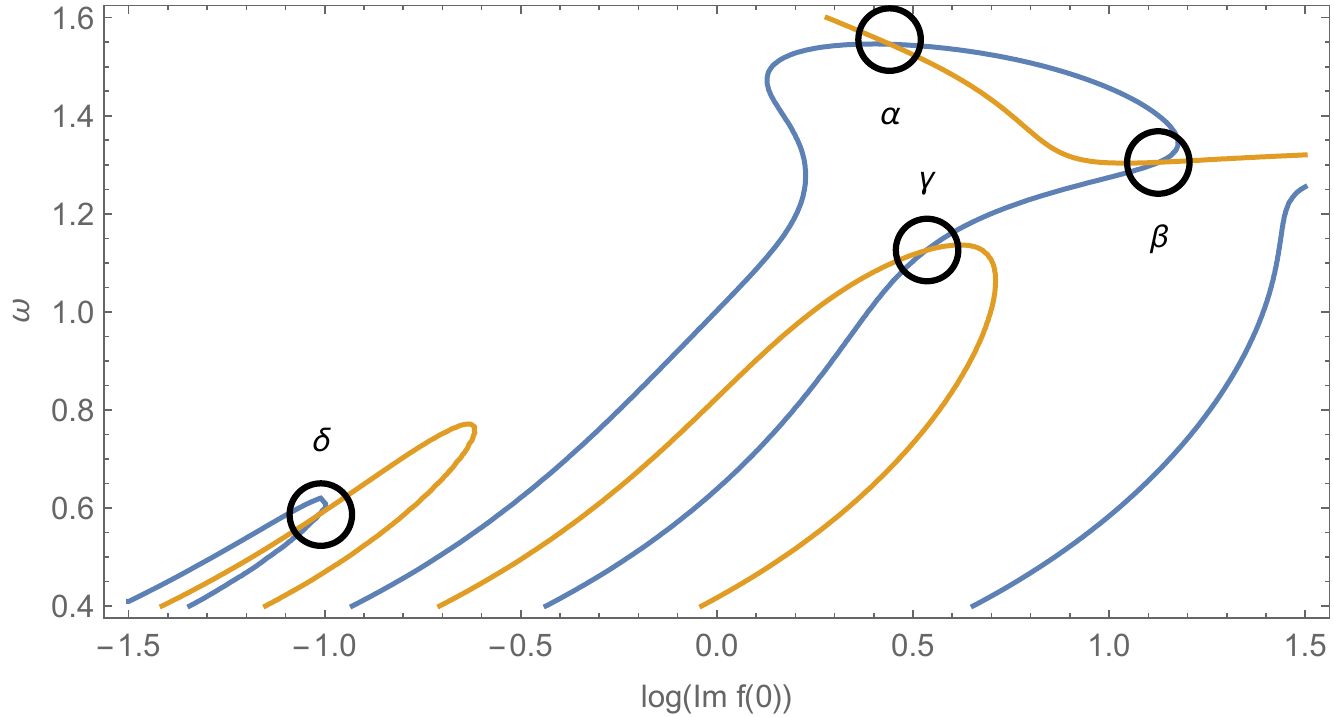}
    \caption{Curves of vanishing real and imaginary parts of $G(\omega,\Im f(0))$ in the $d=5$ hyperbolic case. We can identify four solutions $\alpha$, $\beta$, $\gamma$, $\delta$ in order of decreasing $\omega$.}
    \label{fig:5dhyperbolic}
\end{figure}

We are able in this case to identify reliably four solutions $\alpha$, $\beta$, $\gamma$, $\delta$, which are visible in figure~\ref{fig:5dhyperbolic}. As in four dimensions we cannot exclude the existence of additional solutions for smaller values of $\Im f(0)$, since numerical noise grows significantly in that region.

\begin{center}
\bgroup
\def\arraystretch{2}
\begin{tabular}{|c|c|c|c|}\hline
    Solution & $w$ & $\Im f(0)$ & $z_+$ \\ \hline
    $\alpha$ & $1.546$ & $1.555$ & $1.254$\\
    $\beta$ & $1.305$ & $3.086$ & $1.129$\\
    $\gamma$ & $1.125$ & $1.705$ & $1.109$\\
    $\delta$ & $0.588$ & $0.364$ & $1.156$\\
    \hline
\end{tabular}
\egroup
\end{center}

\section{Conclusions}

We have examined spherically symmetric self-similar gravitational collapses in the Einstein-axion-dilaton system in arbitrary spacetime dimensions $d\geq 4$, and have proposed a practical procedure for finding solutions numerically. We tested the latter in four and five dimensions in the elliptic, parabolic and hyperbolic conjugacy classes and have identified the corresponding solution spaces.

The present numerical procedure for determining scale-invariant solutions represents a significant improvement with respect to previous settings~\cite{Hirschmann_1997}. This is partly due to the algebraic simplifications attained in our starting point, as described in section~\ref{sec:elminateu}. However, we also benefited from improvements in the current version of Mathematica, which allow accurate event detection making use of correspondingly adaptive meshings. These tools simplified considerably the detection of the $z_+$ crossing, avoiding in turn the need for the time-consuming double integration techniques previously employed in previous works, following~\cite{Hirschmann_1997}, which entailed searches over much larger parameter spaces~\cite{hatefialvarez1108,hatefialvarez1307}.

In a subsequent publication~\cite{Antonelli:2019dbi} we shall establish a formalism for the perturbation theory of such self-similar solutions for all dimensions and for all three conjugacy classes, and we shall investigate numerically perturbations of the solutions found in the present paper.\footnote{Note that, some perturbed solutions over spherical symmetric background with small rotating parameter have already been found in a specific model~\cite{Ghodsi_2010}.} One perturbs the generic field $h(t,r)$ of the self-similar solution letting
\begin{equation}
    h(t,r) = (-t)^{\Delta} \left( h_{\text{CSS}}(z) + \varepsilon \,(-t)^{-\kappa} h_1(z) \right)\,,
\end{equation}
where $\Delta$ is the scaling dimension of the field $h$. One then looks for for solutions for the (generically complex) exponent $\kappa$, which define the available modes.


This opens up a channel for computing the Choptuik critical exponent $\gamma$: the exponent of the most relevant mode is related to $\gamma$ through~\cite{KHA}
\begin{equation}
   \frac{1}{\Re\kappa} = \gamma\,.
\end{equation}
We will therefore reconstruct the Choptuik exponent for the elliptic four-dimensional case, which is already known in the literature, and then generalize the analysis to the families of CSS solutions\footnote {Various critical solutions for a massless scalar field have been obtained in ~\cite{Sorkin_2005,Bland_2005}.} in different dimensions and conjugacy classes.

 \section*{Acknowledgments}
We would like to thank Eric Hirschmann, Luis \'Alvarez-Gaum\'e and Augusto Sagnotti for their insights and also for the valuable physics discussions. We are especially grateful to E. Hirschmann for providing various details on the numerical computations and for his physical remarks, and to A. Sagnotti for critical comments on the manuscript.

This work was supported in part by INFN (ISCSN4-GSS-PI), by Scuola Normale, and by the MIUR-PRIN contract 2017CC72MK\_003.

\end{document}